\documentclass[sigconf]{acmart}

\usepackage[utf8]{inputenc} % allow utf-8 input
\usepackage[T1]{fontenc}    % use 8-bit T1 fonts
\usepackage{hyperref}       % hyperlinks
\usepackage{url}            % simple URL typesetting
\usepackage{booktabs}       % professional-quality tables
\usepackage{amsfonts}       % blackboard math symbols
\usepackage{nicefrac}       % compact symbols for 1/2, etc.
\usepackage{microtype}      % microtypography
\usepackage{xcolor}         % colors
\usepackage{graphicx}
\usepackage[]{algorithmic} 
\usepackage{booktabs}
\usepackage{multirow}
\usepackage[ruled,linesnumbered]{algorithm2e}
\begin{document}

\title{Token Factory: Efficiently Integrating Diverse Signals into Large Recommendation Models}

\author{Xilun Chen}
\authornote{Equal contribution. Alphabetical order.}
\email{xilunchen@google.com}
\affiliation{%
  \institution{Google}
  \country{USA}
}

\author{Shao-Chuan Wang}
\authornotemark[1]
\email{scwang@google.com}
\affiliation{%
  \institution{Google}
  \country{USA}
}

\author{Baykal Cakici}
\email{baykal@google.com}
\affiliation{%
  \institution{Google}
  \country{USA}
}

\author{Lukasz Heldt}
\email{heldt@google.com}
\affiliation{%
  \institution{Google}
  \country{USA}
}

\author{Lichan Hong}
\email{lichan@google.com}
\affiliation{%
  \institution{Google}
  \country{USA}
}

\author{Raghu Keshavan}
\email{hkraghunandan@google.com}
\affiliation{%
  \institution{Google}
  \country{USA}
}

\author{Aniruddh Nath}
\email{aniruddhnath@google.com}
\affiliation{%
  \institution{Google}
  \country{USA}
}

\author{Li Wei}
\email{liwei@google.com}
\affiliation{%
  \institution{Google}
  \country{USA}
}

\author{Xinyang Yi}
\email{xinyang@google.com}
\affiliation{%
  \institution{Google}
  \country{USA}
}

\copyrightyear{2026}
\acmYear{2026}
\setcopyright{cc}
\setcctype{by}
\acmConference[RecSys '26]{20th ACM Conference on Recommender Systems}{September 27-October 02, 2026}{Minneapolis, MN, USA}
\acmBooktitle{20th ACM Conference on Recommender Systems (RecSys '26), September 27-October 02, 2026, Minneapolis, MN, USA}
\acmDOI{10.1145/3773078.3831865}
\acmISBN{979-8-4007-2284-4/2026/09}

\begin{abstract}
Large Recommendation Models (LRMs) have demonstrated promising capabilities in industry-scale recommendation tasks. However, holistically integrating traditional signals into these transformer-based architectures effectively and efficiently remains a major challenge. Conventional approaches that "textualize" these signals directly or create discrete item representations often lead to excessively long prompts, substantial memory footprints, and high computational overhead. To overcome these limitations, we propose "Token Factory", a framework designed to transform traditional signals into "soft tokens" that can be directly processed by LRMs. This approach enables efficient integration and compression of heterogeneous input features, preventing prompt length explosion while enhancing model performance. We detail the architecture of Token Factory and present experimental results validating its effectiveness in a production-scale recommendation environment.
\end{abstract}

\maketitle

\section{Introduction}
\begin{figure}[htbp]
\centering
\includegraphics[width=\linewidth]{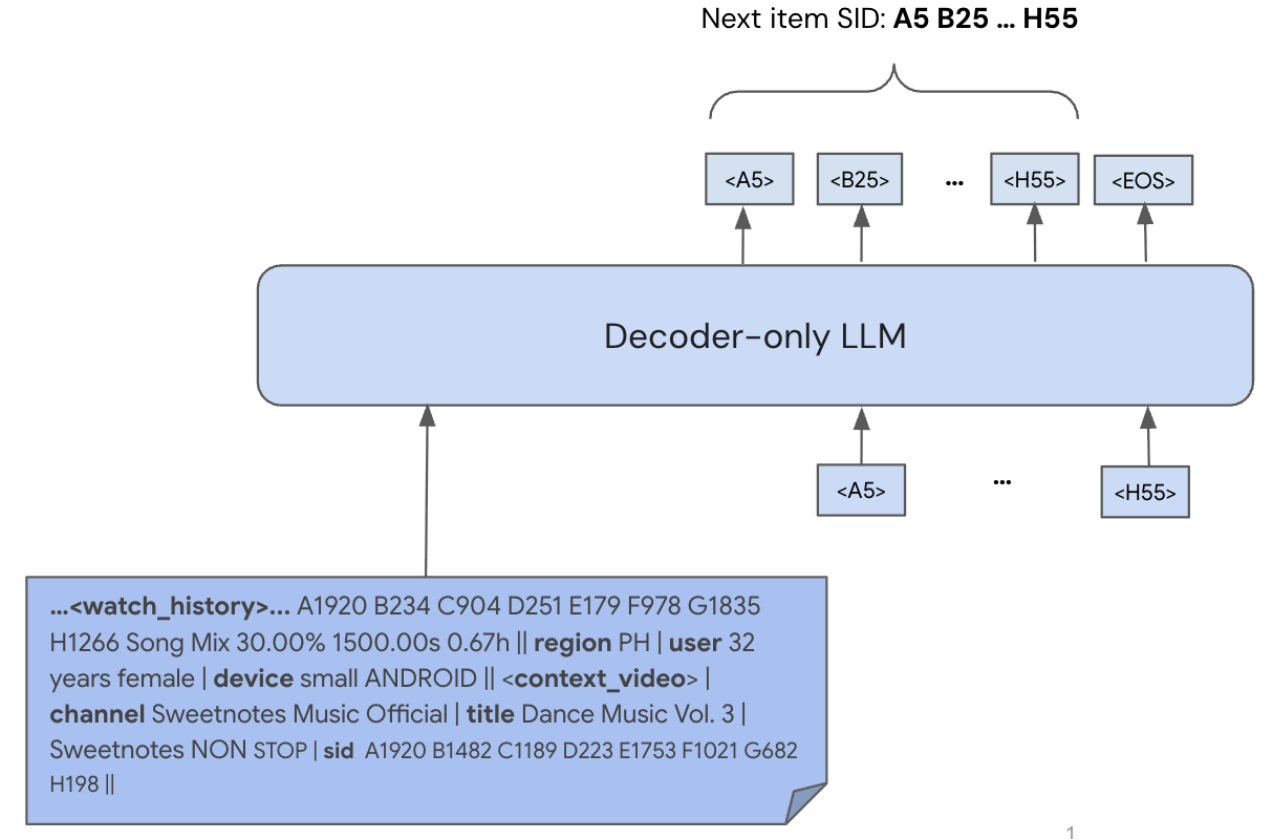}
\caption{Illustration of Generative Retrieval for next video
recommendation from ~\cite{he2025plum}. The input prompt is a sequence of interleaved SID tokens, text, and custom tokens for numerical features.}
\label{fig:next_video_pred}
\end{figure}

In modern industrial recommender systems, Large Recommendation Models (LRMs) based on Transformer architectures have demonstrated great promise~\cite{he2025plum,dai2025onepiece,hu2026ids,huang2025towards}. These models typically rely on a diverse array of signals and advanced processing techniques, including Semantic ID generation~\cite{singh2024better}, custom tokenization, continual pre-training (CPT)~\cite{zhou2024continual}, task-specific fine-tuning~\cite{he2025plum} and user modeling~\cite{tokenminds}. A representative task in this domain is next-video prediction, as illustrated in Figure~\ref{fig:next_video_pred}.

Traditional signals, such as the dense and sparse features that form the foundation of Large Embedding Models (LEMs), carry vital information. They encapsulate direct user interactions and domain-specific insights engineered to effectively guide the recommendation process. However, a primary challenge lies in the efficient representation of these traditional signals within Transformer architectures. Directly mapping these heterogeneous signals to text often results in prohibitive prompt lengths, excessive memory overhead, and high computational costs. For instance, in the next-video prediction task, converting dense signals (such as watch completion ratios and watch times) into text for processing via a custom tokenizer creates a significant computational bottleneck.

To address this challenge, we introduce \textbf{Token Factory}, a framework designed to transform traditional signals into \textit{soft tokens} that serve as an new modality for Large Recommendation Models. A "soft token" is defined as an embedding representation directly within the multi-modal model's embedding space, rather than a discrete ID derived from a fixed vocabulary. Unlike standard "hard" tokens that require a tokenizer to map text to vocabulary indices, soft tokens are generated directly from raw feature values through learnable embedding tables and transformations. These soft tokens are critical because they enable the model to ingest high-dimensional, continuous, and non-textual signals as a new modality without inflating the prompt length, thereby preserving precise information that might be lost or inefficiently represented during conversion to text.

Our core contributions are summarized as follows:
\begin{itemize}
    \item We demonstrate the critical importance of leveraging a rich set of traditional signals in modern Transformer-based Large Recommendation Models (LRMs), such as PLUM \cite{he2025plum}, through extensive experimental analysis, thereby addressing the limitations of current state-of-the-art methods that often restrict these signals due to computational trade-offs.
    \item We introduce \textbf{Token Factory}, a novel architectural framework that enables the effective and efficient mapping of heterogeneous traditional features into "soft token" representations, serving as a new modality for recommendation models.
    \item We validate the empirical effectiveness of Token Factory through offline and online evaluations, including ablation studies, demonstrating that it achieves on-par or superior quality in ranking and retrieval while enabling significant efficiency improvements.

\end{itemize}

The remainder of this paper is organized as follows. We will first discuss related work and present the proposed Token Factory framework. We then share our experimental results and future work.

\section{Related Work}

\textbf{Large Recommendation Models.}
Transformer-based Large Recommendation Models have gained tremendous attention and success in the industry over the past years. Extensive research has focused on effective input representation and model architecture optimization. For example, HSTU~\cite{zhai2024hstu} scales via architectural enhancements and optimization; PLUM~\cite{he2025plum} focuses on leveraging Semantic IDs to represent high-dimensional items for the generative recommendation task; and SEATER~\cite{si2024generative} proposes generating item identifiers that capture the intrinsic semantics of items as indices, rather than treating them merely as text inputs. TokenRec~\cite{qu2025tokenrec} proposed a novel ID tokenization strategy that quantizes masked user/item representations into discrete tokens to incorporate high-order collaborative knowledge. While these approaches excel at modeling user sequences or optimizing architectures, they still rely on textualizing auxiliary features. In contrast, Token Factory provides a unified framework to map heterogeneous signals directly into the Transformer's embedding space as soft tokens, maintaining prompt compactness while preserving feature richness.

The concept of using soft tokens in LRMs is related to CoLLM ~\cite{collm}, which treats collaborative features as a distinct modality and maps pre-trained embeddings into the LLM’s token space via an MLP. While CoLLM is restricted to user and item collaborative embeddings, Token Factory generalizes this approach to accommodate a highly heterogeneous mix of dense, sparse, and sequential signals. Other recent frameworks, such as LLM4Rec~\cite{llm4rec}, integrate diverse modalities (e.g., text, video, and audio) using specialized encoders and cross-modal attention. While powerful, such heavy fusion architectures typically introduce significant computational complexity and prompt-length overhead. Conversely, Token Factory was designed with strict efficiency constraints in mind to ensure compatibility with latency-sensitive industrial environments.

\textbf{Token Compression.}
To further mitigate the scaling challenges when considering long user interaction history, several methods have been proposed to "shorten" or "compress" the input tokens. For example, LONGER~\cite{chai2025longer} optimized representations by introducing a global token and a token merge strategy to merge every $K$ items. GenRec~\cite{zou2026genrec} used a linear projection to encode long user behavior sequences with multi-token item representations based on semantic IDs; however, it does not consider the other features associated with each item, such as user-item interactions. HyMiRec~\cite{zhou2025hymirec} converted items (title + content) into “SID” like codes (3 per item) to reduce input token lengths. Other approaches include segmenting sequences into chunks with learnable tokens and compressing historical segments into personalized expert tokens~\cite{zhang2025efficient}, or using hierarchical K-means clustering~\cite{zhou2025onerec}. Existing compression techniques primarily focus on reducing the number of item IDs in a sequence, often ignoring the rich interaction features associated with each item and solely depending on SIDs or item embeddings. Token Factory allows these features to be fused into the item's soft token representation during compression, ensuring that the compressed representation retains critical information.

\section{Token Factory}
In this section, we present our proposed Token Factory in greater detail.

\subsection{Overview}

\begin{figure}[htbp]
\centering
\includegraphics[width=\linewidth]{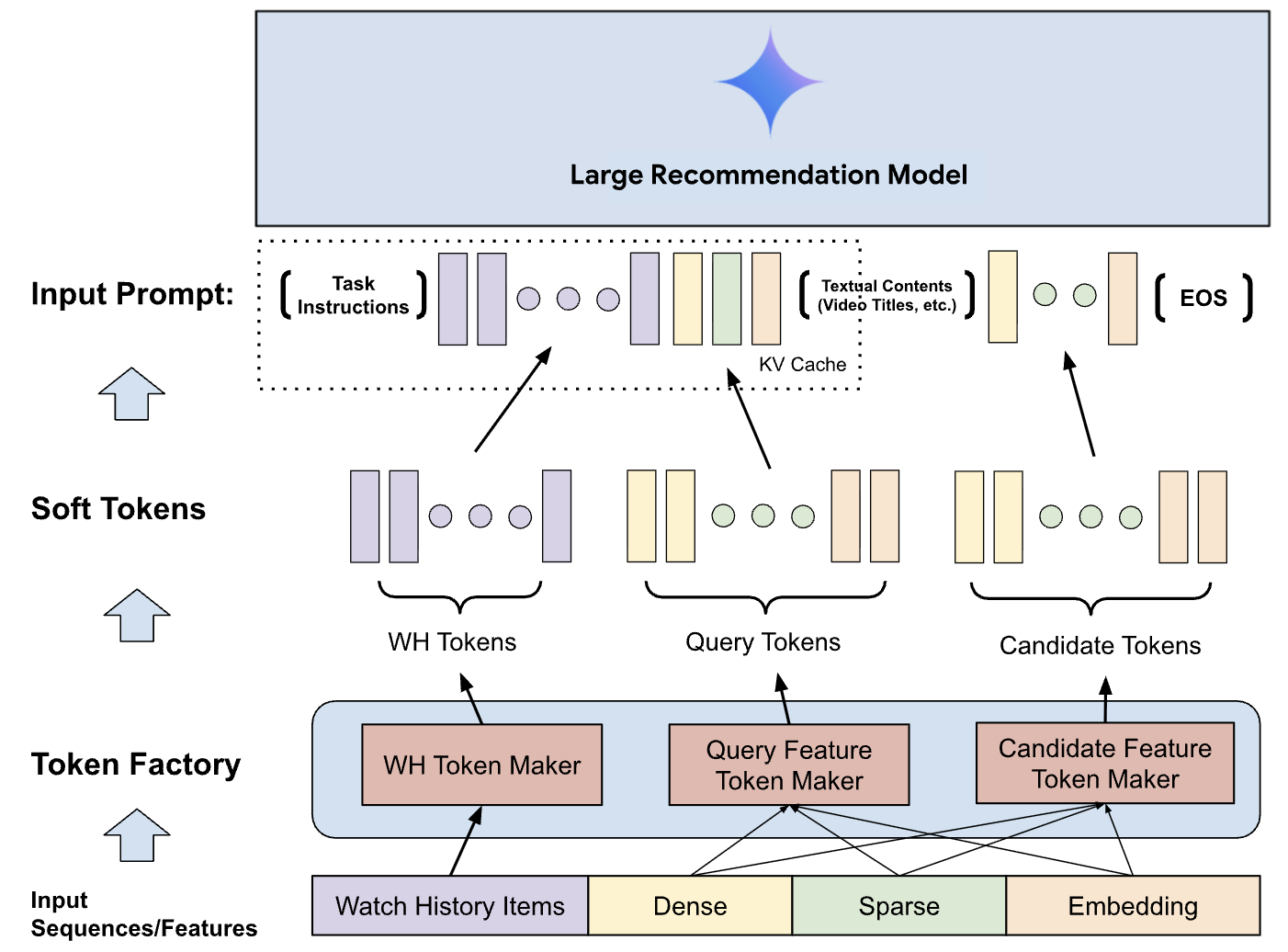}
\caption{Token Factory Architecture. Traditional signals are transformed into "soft tokens" as another modality and integrated with textual tokens in the prompt.}
\label{fig:architecture}
\end{figure}

Figure \ref{fig:architecture} illustrates the overall architecture of Token Factory. Token Factory consists of a collection of \textbf{Token Makers}, each of which produces a sequence of "soft tokens". The traditional signals such as dense, sparse, and embedding features are fed into various token makers and the produced "soft tokens" are then placed at the desired positions in the prompt for the LRM, as shown in Figure \ref{fig:architecture}. 

Specifically, within a video recommendation system scenario, the exemplary Token Factory architecture consists of three token makers, namely 
\begin{itemize}
    \item \textbf{WH Token Maker}, which is responsible for producing soft tokens from the user interaction sequence (e.g., videos in the user watch history).
    \item \textbf{Query Token Maker}, which is responsible for leveraging and integrating query-level features (user features).
    \item \textbf{Candidate Token Maker}, which is responsible for leveraging and integrating candidate-level features (video features or any non-query level features).
\end{itemize}
Notably, these soft tokens can be seamlessly interleaved with textual tokens and benefit from prefix caching. Because query- and user-level features remain static while scoring multiple candidate items within the same recommendation request, their corresponding soft tokens can be pre-computed and cached to significantly reduce inference latency.

This architecture offers several key benefits:
\begin{itemize}
    \item \textbf{Prompt Compactness}: The prompt length can be significantly reduced compared to directly "textualizing" the traditional signals via custom tokenization.
    \item \textbf{Deterministic Length}: The prompt length can be deterministically capped at a predefined budget, irrespective of the number of input features. This is achieved because each Token Maker uses a fixed-output projection layer to map an arbitrary number of concatenated features into a strict, predetermined number of soft tokens.
    \item \textbf{Longer User Interaction}: It enables the consideration of much longer user interaction sequences through compression of multiple signals into one (or fewer) soft token.
    \item \textbf{Efficiency}: It facilitates more efficient training and serving due to the compacted prompt.
\end{itemize}

To realize these architectural benefits, the core component responsible for feature transformation is the Token Maker. In the following section, we provide a detailed exposition of its design and operation.

\subsection{Token Maker}\label{sec:token_maker}
A Token Maker consists of a set of input features and a specification for the target output token(s), encapsulating the logic required to transform these features into the desired number of soft tokens.

To be more specific, let's assume \(F_{input} = [f_1; f_2; ...; f_n]\), where $f_i$ indicates the i-th raw input feature to the Token Maker. Based on the characteristics of the feature, further normalization or embedding lookup could be applied. Therefore, the final features passed to the Token Maker are the concatenation of all transformed features as \(E_{input} = Concat(t_1(f_1), t_2(f_2), ..., t_n(f_n))\), where $t\_i$ is the transformation function to use for feature $f_i$.

We then define a function that further transforms \(E_{input}\) into soft token(s) by

\[T_{output} = G(E_{input})\]

where $T_{output}$ is the $N$ output soft token(s), each of dimension $d_{model}$, which makes them suitable to be directly consumed as embeddings by the Transformer. The function $G$ is a differentiable mapping that transforms $E_{input}$ into an $N \times d_{model}$ embedding vector, which is ultimately reshaped into $N$ soft tokens. Crucially, $G$ is co-trained end-to-end with the Large Recommendation Model, ensuring that the generated soft tokens are aligned with the LLM's semantic space and task objective. The function $G$ can be as simple as a Multi-Layer Perceptron (MLP), or as complex as a neural network or a transformer.

The choice of $N$ balances feature capacity against computational latency. A larger $N$ preserves more high-dimensional information but increases the downstream Transformer’s sequence length and attention overhead. In practice, $N$ is determined empirically based on the feature type: simple features are mapped to a single token ($N=1$), while long interaction sequences are compressed into a small, fixed budget (e.g., $N=10$) to optimize serving efficiency.

\begin{figure}[htbp]
    \centering
    \includegraphics[width=\linewidth]{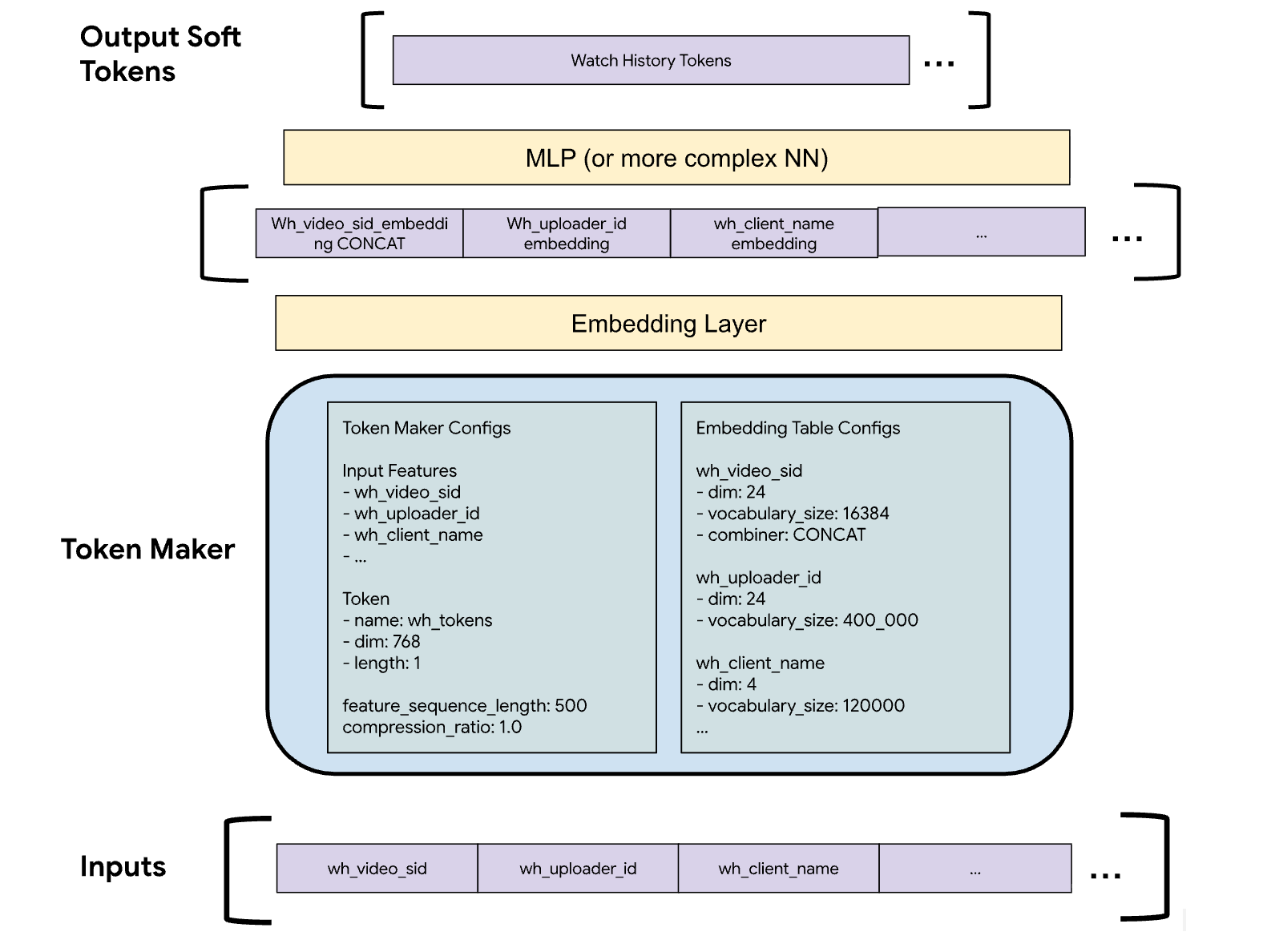}
    \caption{Token maker for user watch history.}
    \label{fig:token_maker}
\end{figure}

Figure~\ref{fig:token_maker} illustrates the core architecture of the Token Maker, which projects all features associated with a single watch history item into a unified soft token representation. In this example, an individual watch item comprises heterogeneous features—including the video Semantic ID (SID), channel name, client information, and watch duration. These features are concatenated and processed through a Multi-Layer Perceptron (MLP) layer to map them into the final soft token embedding space. Because the user history is inherently sequential, sequentially applying this projection transforms the entire watch history into a corresponding sequence of soft tokens within the model input.

Due to the flexibility and extensibility of the Token Maker module, one can naturally leverage ideas from many state-of-the-art token generation approaches, such as \cite{zhu2025rankmixer,jiang2026tokenmixer}.

\subsection{Evolution of Prompts}
To visualize the changes in the prompt, we compare the prompt representations before and after applying Token Factory.

\textbf{Textual Semantic ID Prompt}: In previous work such as PLUM, features like SID were textualized. For example, a user's watch history item might be represented by its SID and associated metadata. An 8-token SID would consume 8 tokens in the prompt. A typical prompt template used in Figure \ref{fig:next_video_pred} might look like:

\noindent\fbox{%
  \begin{minipage}{\linewidth}
    [watch\_history] | region [user\_region] | user [user\_age] [user\_gender] device [device\_info] | video [video\_title]...
  \end{minipage}%
}
The actual prompt would expand to:

\noindent\fbox{%
  \begin{minipage}{\linewidth}
    A100 B100 C100 D100 E100 F100 G100 H100 5.00\% 50.00s 2.5h | ... region PH | user 32 years female | device small ANDROID | video baby shark...
  \end{minipage}%
}
where $A100$ to $H100$ are the 8 tokens for SID, together with a few dense features, i.e., watch completion ratio, video watch time, and time since watch, that are repeated across the length of the actual watch history. This results in very long prompts for even moderate watch history lengths.

\textbf{Token Factory Prompt}: With Token Factory, the traditional features (dense, sparse, sequence) are transformed into soft tokens. The prompt length is significantly reduced. The soft tokens (indicated by $<emb>$ below) do not have a human-readable text representation as they are embedding vectors, but conceptually the prompt would look like this:

\noindent\fbox{%
  \begin{minipage}{\linewidth}
    [watch\_history]<emb><emb><emb>...<emb>[end\_watch\_history] <emb>...<emb> video baby shark <emb><emb>...<emb>
  \end{minipage}%
}

In this example, each $<emb>$ token between [watch\_history] and [end\_watch\_history] corresponds to a single watch history item, replacing the verbose textual representation (e.g., 'A100 ... 2.5h'). The number of these $<emb>$ tokens matches the history length. We retain some textual tokens, such as the video title, to leverage the LLM's natural language understanding capability. The additional $<emb>$ tokens following the title are generated from other heterogeneous features of the user, candidate item, and their interactions.

With this transition, items are no longer represented by discrete sequences of textual SID tokens. Consequently, the individual sub-tokens associated with different watch history items do not participate in direct, token-level attention. To verify that the model still successfully captures interactions across watch history items despite the absence of explicit SID tokens, we present a comparative analysis of the attention weights for both soft tokens and textual SIDs in Appendix A. This analysis demonstrates that the compressed soft tokens continue to facilitate sufficient and effective cross-item attention within the interaction sequence.

\subsection{Sequence Compression} \label{sec:compression}
In Section~\ref{sec:token_maker}, we showed an example of how one watch history item transforms into one soft token via the Token Maker. However, this may not be sufficiently efficient when dealing with longer watch histories (e.g., if we want to encode thousands of watches or even a user's lifetime watches). Therefore, we need to further compress the sequence into fewer soft tokens.

\subsubsection{MLP Compression}
One approach to further compression is using an MLP on the sequence dimension. Assume that the Token Maker has already converted the watch history to soft tokens, where one item is represented by one token. This results in a JAX~\cite{jax2018github} array of shape $[batch\_size, sequence\_length, token\_dim]$, where $sequence\_length = N$. An MLP layer of shape $[N, \dots, M]$ can be applied on the second dimension to reduce the sequence length from $N$ to $M$. As a result, we end up with $M$ soft tokens representing the entire sequence, achieving a compression ratio of $M / N$.

\subsubsection{Attention Pooling}
Inspired by~\cite{chai2025longer}, we could also apply a lightweight transformer and use attention pooling on the last item in every $K$ items in the sequence. This essentially summarizes every $K$ items using one soft token and results in a compression ratio of $1 / K$.

\section{Experiments}
In this section, we conduct a comprehensive set of experiments on various use cases within our Large Recommendation Model PLUM framework, including ranking and retrieval tasks, to validate the proposed Token Factory framework. We first evaluate the performance of the PLUM-based ranking model and the PLUM-based generative retrieval model to demonstrate the efficiency and effectiveness of Token Factory. We then perform various ablation studies to understand the importance of traditional signals and Semantic IDs that are used in Token Factory, as well as the impacts of user interaction history length.

\subsection{Results on Ranking Task} \label{sec:plum_ranker_results}
In this section, we detail our experimental results for PLUM with Token Factory in the ranking task (i.e., predicting the CTR of a given video).

\subsubsection{Experiment Setup}
\label{sec:plum_ranker_exp_setup}
We run the experiment on a PLUM model derived from 110M Mixture-of-Exports(MoE) variant of Gemini encoder model. Training data consists of user watch history, traditional dense and sparse features, and the clicks for the next video prediction task.
\begin{itemize}
    \item \textbf{Baseline}: The baseline uses the textual Semantic ID and custom tokenization of several dense signals using the same PLUM model. In the baseline prompt, each watch history item takes 12 tokens (8 tokens for the SID, 1 token for the channel name, and 3 tokens for dense features in textual format). We use a prompt length of 1536 tokens as input, which contains 200 watch history items, user demographic information, and the candidate video title in textual format. Excessive tokens beyond the 1536 limit are truncated from the left, which specifically discards the oldest watch history items to preserve the user's most recent interactions and target metadata.
    \item \textbf{Treatment}: The treatment uses Token Factory to transform each watch history item into one soft token. As a result, the prompt length is significantly reduced to 480 tokens.
\end{itemize}

\subsubsection{Experiment Results}
\begin{figure}[htbp]
    \centering
    \includegraphics[width=\linewidth]{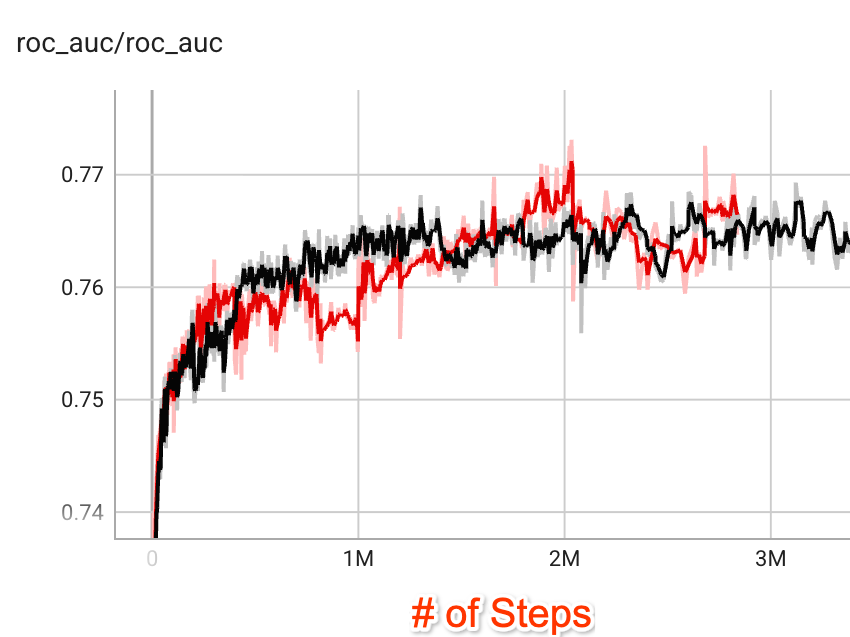}
    \caption{ROC AUC comparison between baseline (black line) and treatment (red line), keeping the same batch size.}
    \label{fig:plum_ranker_e1}
\end{figure}

The primary research question we aim to answer in this evaluation is: \textit{Can Token Factory achieve on-par predictive quality compared to the baseline while operating on a significantly compressed prompt?} Figure~\ref{fig:plum_ranker_e1} addresses this by showing the comparison between the baseline and the treatment on ROC AUC for the CTR prediction task. Both models use the same global batch size even though the treatment has a much shorter input length. We observe that the model using Token Factory that converts textual watch items into soft tokens tends to have a lower AUC initially, but catches up to the baseline at 1.5M steps and consistently performs comparably afterwards. This initial lag is expected because Token Factory introduces additional randomly initialized parameters in the token makers and new embedding tables for sparse features. These components must learn from scratch how to effectively project the raw signals into the LLM's semantic space. In contrast, the baseline maps features to text, which may be easier for the model to optimize initially, albeit less efficiently. Once the token makers are sufficiently trained, the treatment catches up and matches the baseline's quality, demonstrating that the soft token representation is as effective as the full textual representation while being much more compact.

With this setup, we can also achieve a significant improvement in training speed of about 200\% faster due to the fact that the prompt length is only about 30\% of the baseline length.

\begin{figure}[htbp]
    \centering
    \includegraphics[width=\linewidth]{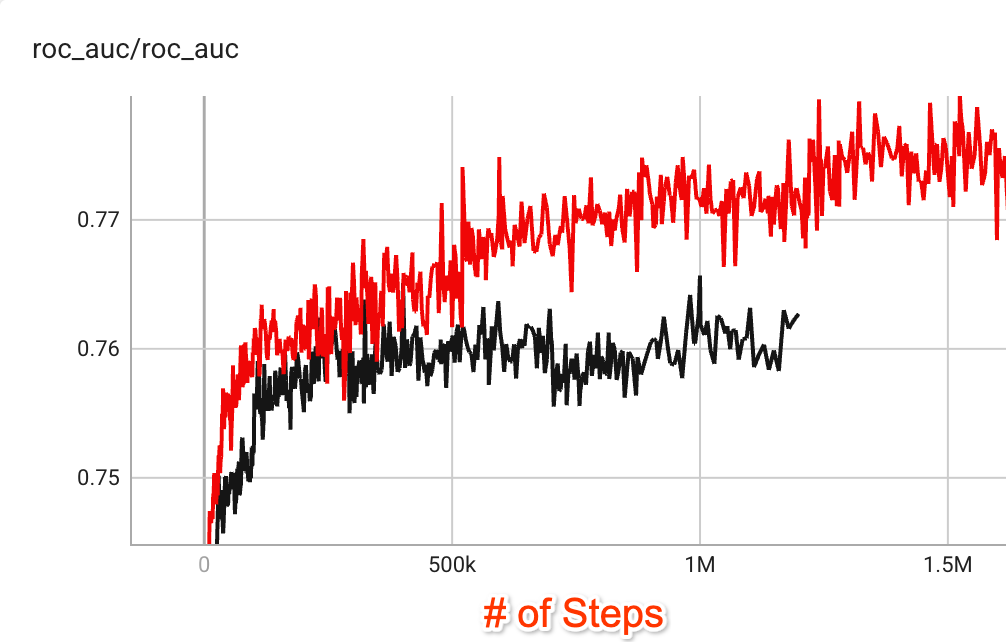}
    \caption{ROC AUC comparison between baseline (black line) and treatment (red line), with increased batch size.}
    \label{fig:plum_ranker_e2}
\end{figure}

Figure \ref{fig:plum_ranker_e2} shows the comparison between the baseline and the treatment on ROC AUC for the CTR prediction task when we increase the global training batch size for the treatment by 200\%. We observe that now the model with Token Factory is outperforming the baseline even at a very early stage due to the increased batch size. The efficiency gain from Token Factory unlocks the consideration of much bigger batch size and improves the overall performance of the model.

\subsection{Results on Generative Retrieval Task}
In this section, we detail our experimental results for PLUM with Token Factory in the generative retrieval task (i.e., predicting the Semantic IDs for the next video to watch).

\subsubsection{Experiment Setup}
We run the experiment on a PLUM model derived from 210M Mixture-of-Exports(MoE) variant of Gemini decoder model. Training data consists of user watch history, traditional dense features, and the semantic IDs for the next video as the target.
\begin{itemize}
    \item \textbf{Baseline}: The baseline relies on textual Semantic IDs and custom vocabulary to represent dense signals. In this setup, each watch history item consumes 5 tokens (1 token for a random hash of the SID sequence representing an item, 1 token for a random hash of the channel name, and 3 tokens for dense features in textual format). The total prompt length is capped at 768 tokens, which includes the user's demographic information in textual format.
    \item \textbf{Treatment}: The treatment uses Token Factory to transform each watch history item into one soft token. The prompt length is 256, which includes 200 watch history items. The user demographic features used in textual format in the baseline are also converted into soft tokens.
\end{itemize}

\subsubsection{Experiment Results}
We use recall@10 to measure model's offline performance and also report our online experiment results in production. For online performance, we use "Unique Impressions," which measures the candidates that are uniquely generated by the model and shown to the user, and Satisfied Watchers, and Satisfied Watch Time to measure user engagement of the new model.

Table \ref{tab:plum_retrieval_home} presents both offline and online evaluation results for the generative retrieval task on YouTube Homepage recommendation. As shown in the table, the model using Token Factory achieves a +2.0\% improvement in offline Recall@10 compared to the baseline. In online experiments, we observed a significant improvement on Unique Impressions (increased by +16.8\%), and notably, Unique Impressions for one-day-fresh videos soared by 67.1\%. This demonstrates that Token Factory enables the LRM to effectively leverage holistic signals from features and efficiently consider longer user interaction sequence, greatly enhancing its ability to retrieve and recommend fresh content. Furthermore, these improvements in exploration did not come at the cost of user satisfaction; we achieved positive gains in both Satisfied Watchers (+0.04\%) and Satisfied Watch Time (+0.05\%).

\begin{table}
    \centering
    \begin{tabular}{|c|p{1.2cm}|p{1.5cm}|p{1.2cm}|p{1.2cm}|}
        \hline
        Offline & \multicolumn{4}{|c|}{Online} \\ \hline
        Recall@10 & Unique Impressions & Unique Impressions (1d fresh) & Satisfied Watchers & Satisfied Watch Time \\ \hline
        +2.0\% & +16.8\% & +67.1\% & +0.04\% & +0.05\% \\ \hline
    \end{tabular}
    \caption{Offline and Live Experiments for using Token Factory in a PLUM based model in the retrieval stage for YouTube Homepage recommendation. The new model is much more effective for new videos, as measured by the Unique Impressions of one-day-fresh videos.}
    \label{tab:plum_retrieval_home}
\end{table}

\subsection{Scaling Studies}
In this section, we investigate the following research question: \textit{Does Token Factory primarily serve as an efficiency optimization, or does the soft token representation also provide headroom for quality improvements?} To answer this, we designed four ablation studies to isolate the factors impacting soft token performance in the ranking task. All treatments and the baseline use a fixed prompt length of 480 tokens (which contains 200 watch history items, the candidate video SID, the video title, and associated traditional dense and sparse features) with a batch size of 32k. When the actual prompt exceeds this length, it is truncated from the left.

\begin{itemize}
    \item \textbf{Baseline}: Same setup as the treatment in Section \ref{sec:plum_ranker_exp_setup}.
    \item \textbf{WH\_SID}: Similar to Baseline but uses the textual SID format for watch history as oppose to using soft tokens for watch history
    \item \textbf{NO\_FEAT}: Similar to Baseline but removes all traditional dense and sparse features (except those from the watch history).
    \item \textbf{WH\_SID\_NO\_FEAT}: Similar to \textbf{WH\_SID} but removes all traditional dense and sparse features.
    \item \textbf{NO\_FEAT\_STRICT}: Similar to \textbf{NO\_FEAT}, but further removes the dense and sparse features from the watch history as well. With this setup, a soft token for watch history only contains the SID and channel ID.
\end{itemize}

\begin{figure}[htbp]
    \centering
    \includegraphics[width=1\linewidth]{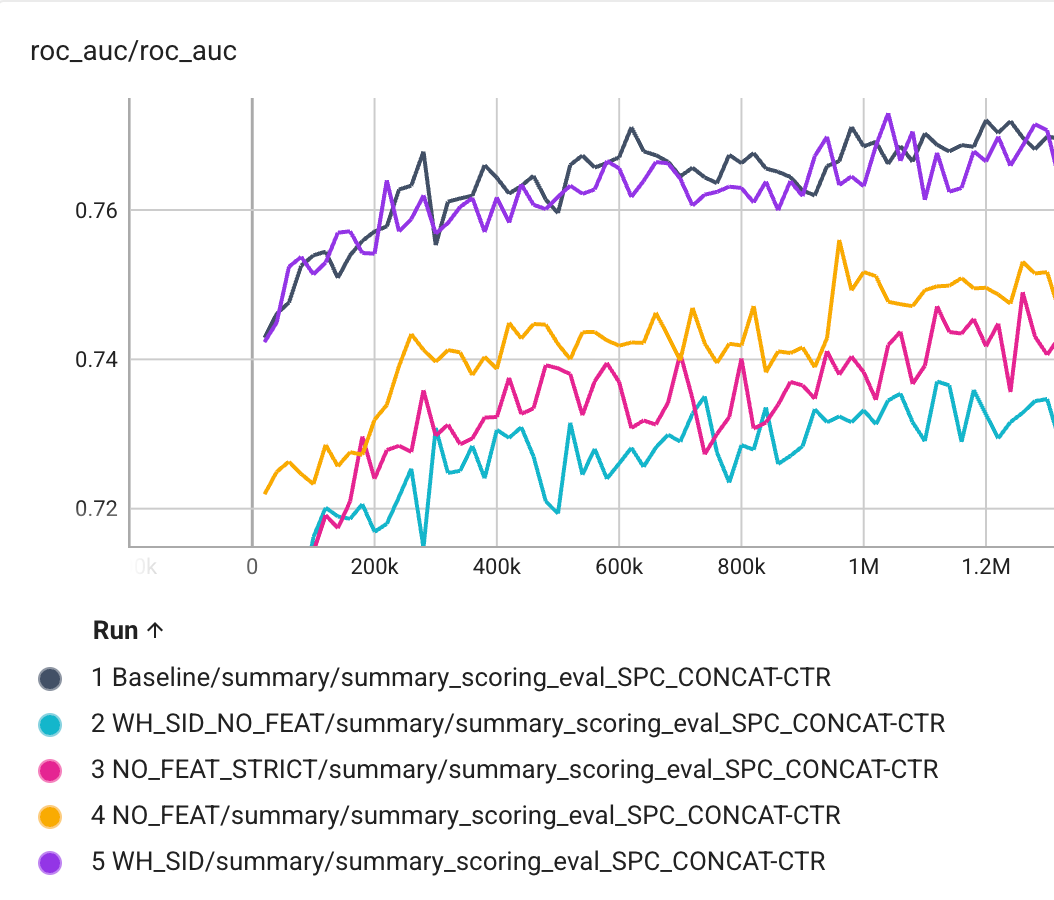}
    \caption{ROC AUC comparison.}
    \label{fig:ablation_study}
\end{figure}

Figure \ref{fig:ablation_study} shows the AUC comparisons and we observe that:
\begin{itemize}
    \item Using soft tokens or the SID format did not show a clear gap when all dense and sparse features are present.
    \item Comparing \textbf{NO\_FEAT\_STRICT} with \textbf{WH\_SID\_NO\_FEAT}, we see that using soft tokens for the watch history performs better than using SIDs in textual format. This is mostly due to the context window budget or constraint (480) that we set: the soft token models were able to include 200 watch history items into the prompt, whereas SIDs may get truncated due to budget constraints.
    \item When comparing \textbf{NO\_FEAT} and \textbf{NO\_FEAT\_STRICT}, we can conclude that adding more features in each soft token still helps with the CTR prediction task.
\end{itemize}

\subsubsection{Longer Watch History}
With the compression capabilities described in Section \ref{sec:compression}, we further conducted a study to extend the watch history from 200 to 500 in the CTR prediction task and use a compression ratio of 10\% to compress the 300 added watch history items into 30 soft tokens. We observe a +0.08\% improvement in AUC. Note that in this study, the treatment has an increased prompt length of 30 tokens compared to baseline to hold the extra compressed watch history soft tokens.

\section{Conclusion}
In this paper, we present Token Factory, a novel framework designed to efficiently integrate diverse traditional signals into modern Transformer-based Large Recommendation Models. By transforming heterogeneous features such as dense, sparse, and sequence signals into "soft tokens," our approach successfully circumvents the challenge of prompt length explosion that arises from direct textualization. Our experimental results on both ranking and generative retrieval tasks in an industrial-scale recommendation system demonstrate that Token Factory not only significantly reduces training and serving costs but also maintains or improves model quality.

\begin{acks}
We would like to thank Min-hsuan Tsai for providing valuable feedback to this paper, and the following individuals for applying Token Factory to various surfaces and use cases (alphabetical order): Dapeng Hong, Vikas Kedigehalli, Yuxuan Li, Sunny Liu, Yang Liu, Steven Su, Yueqi Wang, Bo Yan, Anlan Yu, Qiao Zhang, Rein Zhang, and Zelong Zhao.
\end{acks}

\bibliographystyle{ACM-Reference-Format}
\bibliography{references}

\appendix
\section{Soft Tokens Attention Visualization}
To better understand the behavioral differences between models utilizing soft tokens and those using textual SIDs (similarly as ~\cite{he2025plum}), we analyze and compare their internal attention patterns. Specifically, we visualize the attention distributions across all transformer layers and attention heads for both options in the context of ranking task.

\begin{figure}[htbp]
    \centering
    \includegraphics[width=1\linewidth]{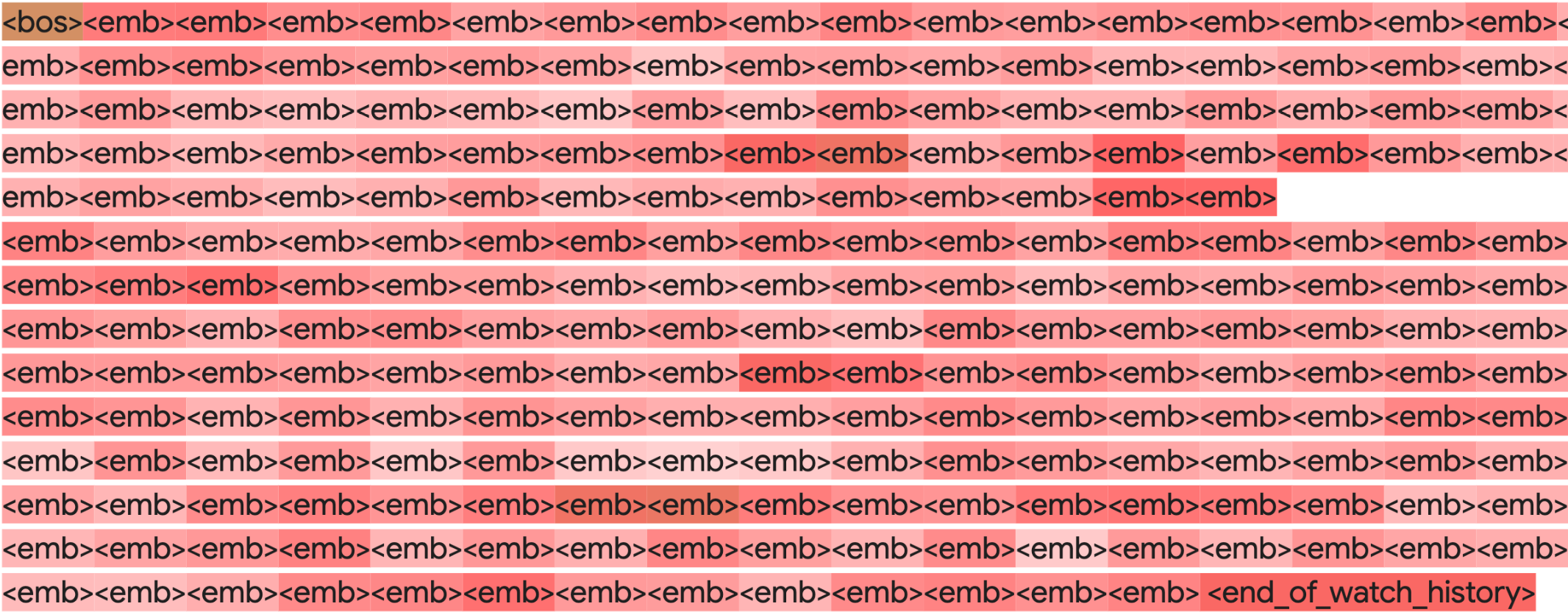}
    \caption{Max attention scores across all layers/heads for Soft token inputs}
    \label{fig:soft_max}
\end{figure}

\begin{figure}[htbp]
    \centering
    \includegraphics[width=1\linewidth]{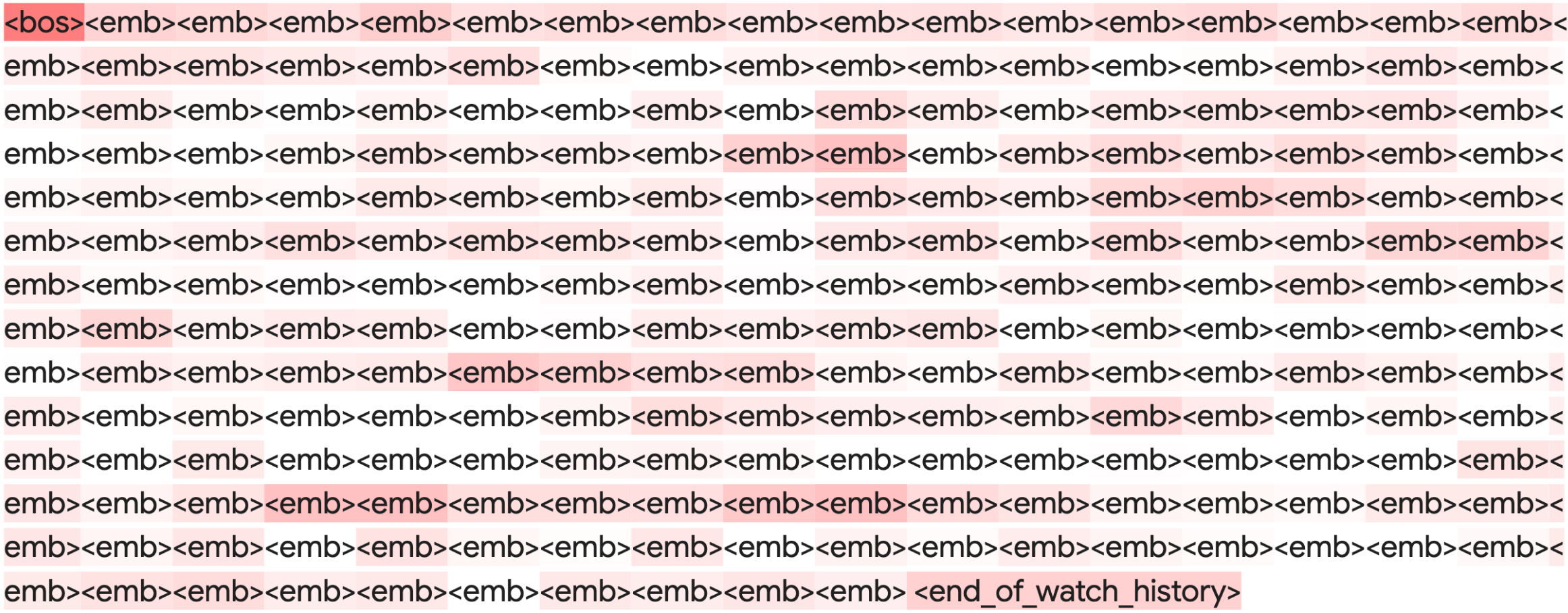}
    \caption{Mean attention scores across all layers/heads for Soft token inputs}
    \label{fig:soft_mean}
\end{figure}

\begin{figure}[htbp]
    \centering
    \includegraphics[width=1\linewidth]{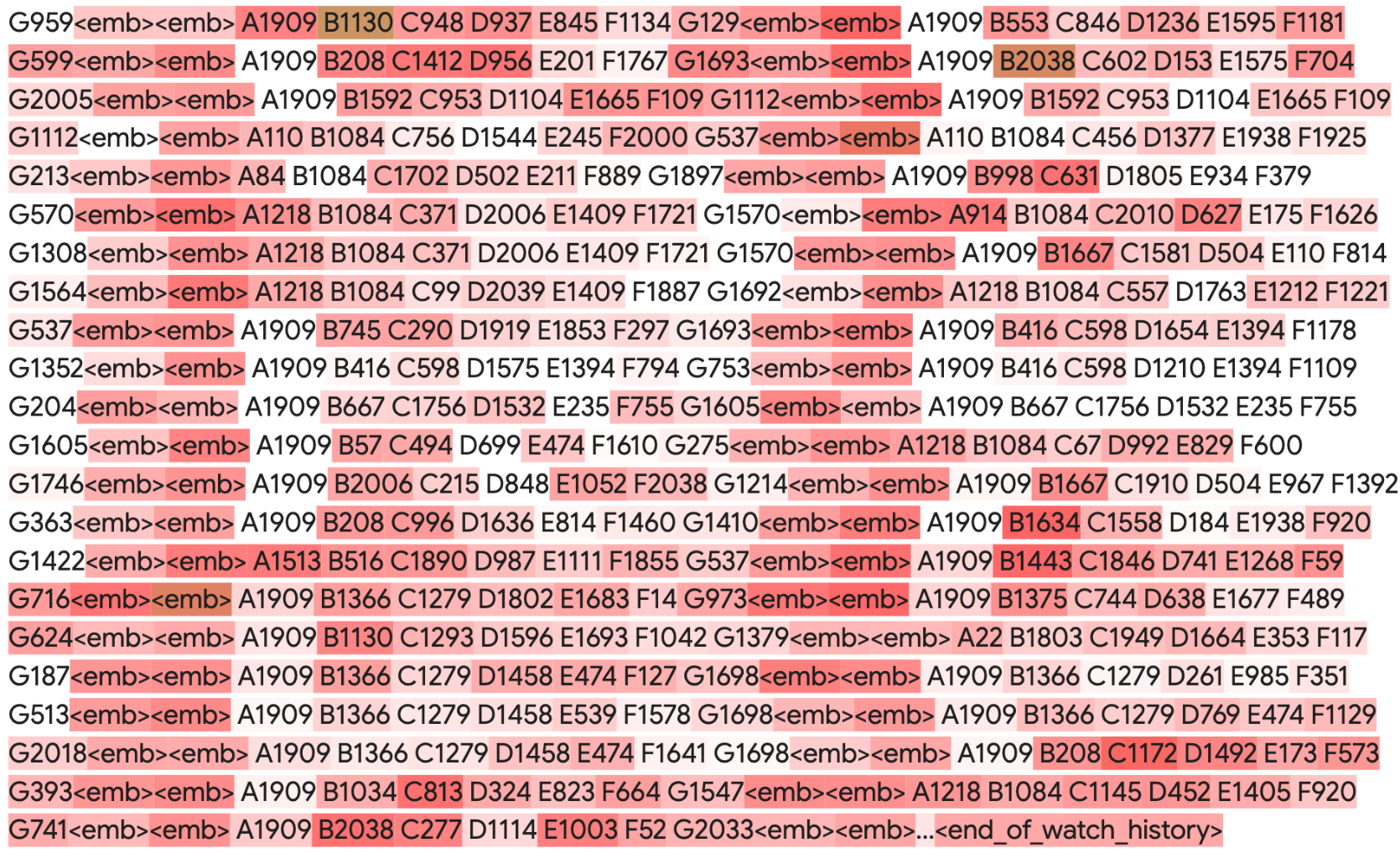}
    \caption{Max attention scores across all layers/heads for textual SID inputs}
    \label{fig:text_sid_max}
\end{figure}

Figures~\ref{fig:soft_max} and~\ref{fig:text_sid_max} present the maximum attention scores observed across all layers and heads for the soft token and textual SID inputs in the user watch history (note that in the textual SID configuration, two \textit{<emb>} tokens are appended to each SID to represent the channel and certain dense features). In these heatmaps, higher attention scores are indicated by higher color intensity. 

In the textual SID model (Figure~\ref{fig:text_sid_max}), we observe that certain frequent tokens, such as \textit{A1909}, which commonly appears as the leading sub-token of the SID sequence receive very little attention across the interaction history. More importantly, nearly half of the textual SID tokens in the prompt are virtually unattended, appearing as inactive (white) regions. In contrast, the soft token model (Figure~\ref{fig:soft_max}) shows a non-trivial level of attention across all soft tokens (visualized as varying shades of red). This indicates that by compressing heterogeneous features into soft tokens, every token is leveraged by at least one specialized sub-network or attention head, resulting in a more efficient and comprehensive utilization of the model's capacity.

\begin{figure}[htbp]
    \centering
    \includegraphics[width=1\linewidth]{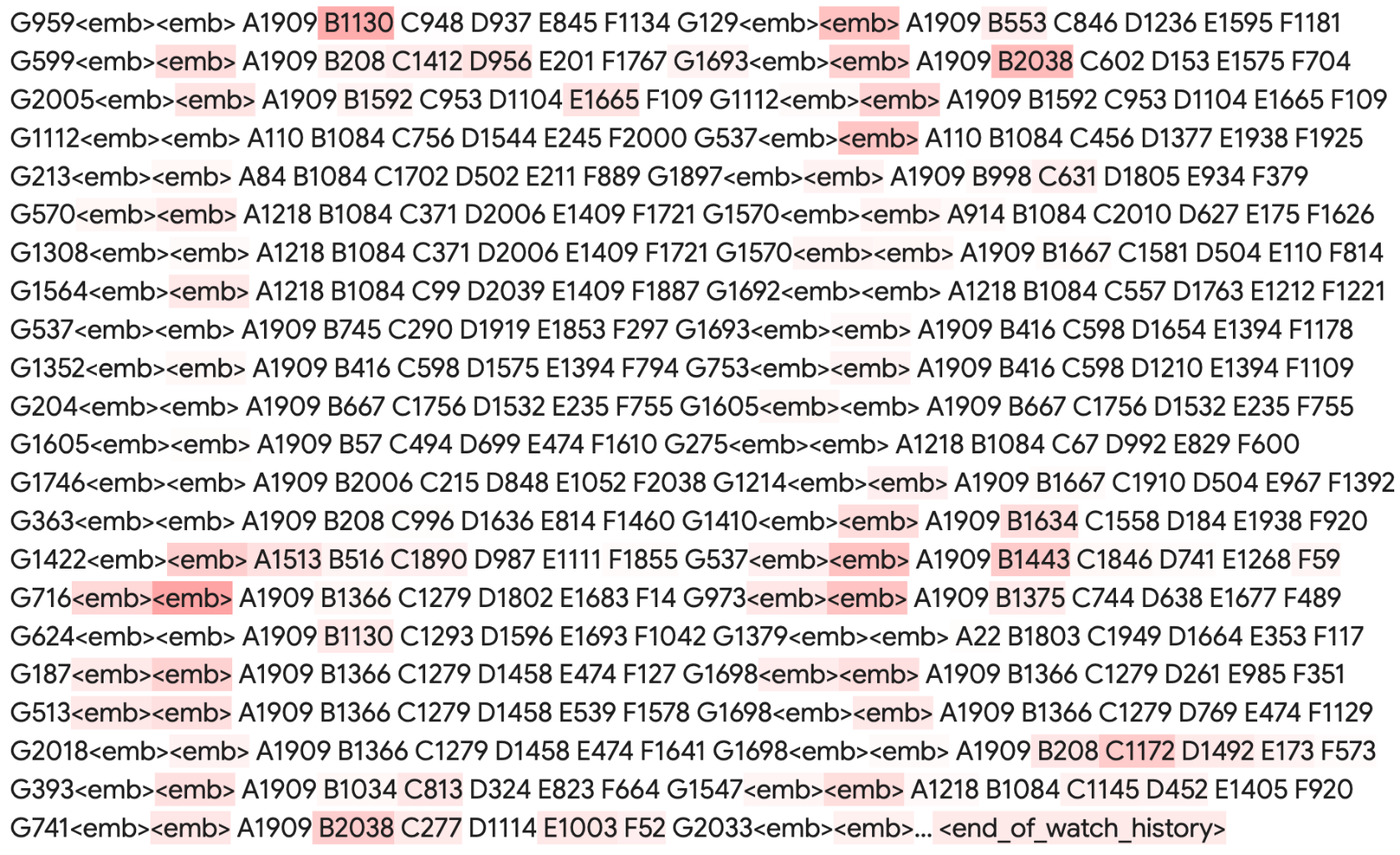}
    \caption{Mean attention scores across all layers/heads for textual SID inputs}
    \label{fig:text_sid_mean}
\end{figure}

Figures~\ref{fig:soft_mean} and~\ref{fig:text_sid_mean} illustrate the mean attention scores averaged across all layers and heads. In the textual SID configuration (Figure~\ref{fig:text_sid_mean}), the attention is highly sparse and concentrated on a very small subset of tokens. This concentration suggests substantial representation redundancy, with the majority of layers and heads focusing on the same few tokens. Conversely, the mean attention map for the soft token model (Figure~\ref{fig:soft_mean}) is significantly denser and more distributed. This indicates that the attention mass is spread across the sequence, allowing different attention heads and layers to capture complementary signals from the input.

In summary, these visual analyses demonstrate that the soft token representations generated by Token Factory are not only computationally efficient due to prompt compression, but they are also utilized more effectively and balanced by the attention mechanism, successfully mitigating attention redundancy.

\end{document}